# Dendritic cell-cluster metasurface manipulation of visible light

Z. H. Fang,[1] H. Chen,[1] D. An,[1] C. R. Luo[1] & X. P. Zhao[1,*]

[1]Smart Materials Laboratory, Department of Applied Physics, Northwestern Polytechnical University, Xi'an 710129, P.R.China.
*Corresponding author: xpzhao@nwpu.edu.cn

**The manipulation of visible light is important in science and technology research. Metasurfaces can enable flexible and effective regulation of the phase, polarization, and propagation modes of an electromagnetic wave. Metasurfaces have become a research hotspot in optics and electromagnetics, and cross-polarization conversion is an important application for visible-light manipulation using a metasurface. A metasurface composed of nano-antenna arrays and bilayer plasma can reportedly convert the direction of linear polarized light efficiently. However, the metasurface of cross-polarization conversion operating in short-wavelength visible light is problematic. In addition, previous metasurfaces prepared using the top–down etching method is unsuitable for practical applications because of the necessary harsh experimental conditions and the high construction cost of preparation. In the present work, we suggest a dendritic cell-cluster metasurface achieve cross-polarization in transmission mode within 550, 570, 590 and 610 nm wavelength. Preparation is accomplished using a bottom-up electrochemical deposition method, which is easy and low cost. The dendritic cell-cluster metasurface is an important step in cross-polarization conversion research and has broad application prospects and development potential.**

http://

Visible light is likely to become the main medium of communication and information processing of the next generation. In recent years, researchers tried to manipulate light through a variety of means. Owing to artificially designing different cell structure, metamaterials achieved many singularities which inexistence in nature, such as negative refraction, anomalous Cerenkov radiation, anomalous Doppler effect, perfect lens, super-resolution imaging, invisibility cloak, and electromagnetic wave polarization rotation capability [1-3]. These singularities are attracting more and more researchers to participate in the research on metamaterials. Following significant research results in microwave wavelength and infrared [4-5], the research on metamaterials operating in visible light became more and more important [6]. As two-dimensional metamaterials, metasurface reserve the singularities of three-dimensional counterparts in manipulating electromagnetic wave behaviors while reducing the challenges in fabrication [7]. Notably, ultrathin metasurfaces have been comportable designed to deflect a propagating light into anomalous refraction channels [8-13], obeying "generalized Snell's laws", by imparting phase discontinuities. The metasurface thickness is far smaller than the operational wavelength, allowing, in principle, the miniaturization and integration of optical components [14]. Lee *et al.* [15] proposed and experimentally realized metasurfaces based on the coupling of electromagnetic modes in plasmonic metasurfaces with quantum-engineered electronic intersubband transitions in semiconductor heterostructures.

Polarization is an important characteristic of light; efficient manipulations over light polarizations are always desirable in practical applications. Several applications are reported for cross-polarization rotation, such as array of nanoantennas, plasmonic, and dielectric [16-27]. Lin *et al.* [17] described the dielectric gradient metasurface optical elements capable of also achieving high efficiencies in transmission mode in the visible spectrum. The cross-polarization rotation has achieved certain results in the frequency range of visible light. It is noteworthy that Qin *et al.* [14] reveal that the cross-polarization conversion efficiency may be further increased to 36.5% with an optimization of the proposed structure in the wavelength of 815 nm. The cross-polarization conversion research has great development space in short-wavelength visible light. Gansel [24] investigated the propagation of light through a uniaxial photonic metamaterial composed of three-dimensional gold helices arranged on a two-dimensional square lattice. These nanostructures are fabricated via an approach based on direct laser writing into a positive-tone photoresist followed by electrochemical deposition of gold. In addition, the majority of proposed metasurfaces is prepared using the top–down mechanical erosion methods which needs harsh experimental conditions and high cost. A new metasurface structure with preparation convenience and light manipulation capability is necessary. The bottom-up fabrication of facile, lower cost, and larger areas of optical metasurface provided important benefits for applications in negative refraction and slowing down light [28-29]. The dendritic metasurface which is prepared using the bottom-up electrochemistry deposit method can realize abnormal reflection and refraction effects in visible light.

Here, the dendritic cell-cluster metasurface achieving effective cross-polarized conversion in transmission mode was numerically and experimentally demonstrated. In this paper, the suggested metasurface, composed of dendritic cell, was prepared with the bottom-up electrochemistry deposit method. It is cost-effective and did not use expensive equipment or harsh conditions. The cross-polarized transmitted light is deflected from normal when passed through the dendritic metasurface. Significant cross-polarization conversion has been achieved in the visible-light wavelengths of 550, 570, 590, and 620 nm. These results represent a significant improvement in visible-light manipulation.

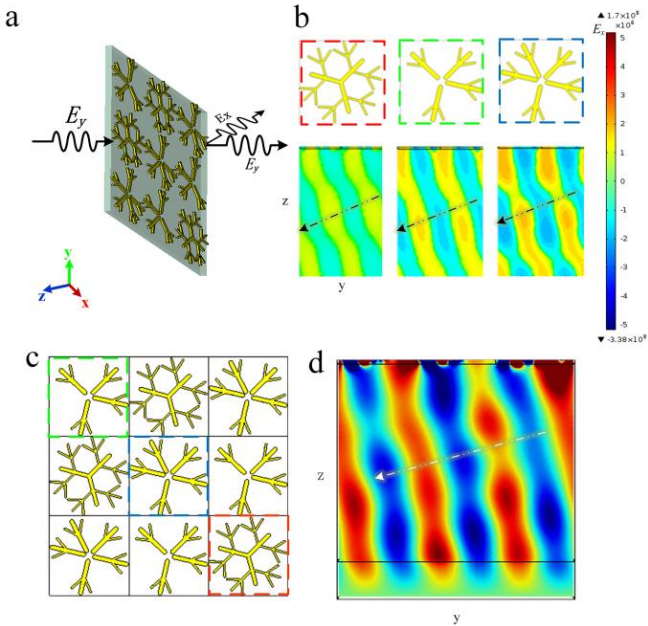

**Fig. 1.** (a) Schematic of the linear polarized plan wave perpendicularly go through the dendritic metasurface; (b) schematic structural view of three types of dendritic cells and the corresponding field distribution of transmission electric field; (c) schematic structural view of a dendritic cell cluster composed of three types of dendritic cells, the same type of dendritic cell is marked in same color dashed box in (b) and (c); (d) the transmission electric of the dendritic cell cluster metasurface.

Linearly polarized light is perpendicularly incident to the dendritic metasurface in CST software simulation. Fig. 1a shows that the samples are horizontally placed in the XY plane. The material on the $SiO_2$ substrate is dendritic structured Ag, whose relative permittivity can be described by the Drude model. The incidence wave in all cases is set as an $E_y$ linearly polarized plane wave perpendicularly transmitted to the surface along −z. Co- and cross-polarized transmitted plane wave $E_y$ and $E_x$ are obtained after the incident linearly polarized plane wave through the dendritic metasurface. The detailed structure of dendritic metasurface is shown in Fig. 1b. Fig. 1b shows three types of metasurface. Each metasurface is composed by a single type of dendritic structure. The electric field in the x direction of the transmitted wave is also shown in Fig. 1b. The dotted arrows indicate the direction of the electric field. An $E_x$-linearly polarized transmitted plane wave is obtained and deflected into the anomalous refraction channel. Three types of dendritic structure are randomly arranged to form a dendritic cell cluster, and the number of each type of dendritic structure exist in a cluster is 3 (Fig. 1c). The different dendritic structure is indicated by a dashed box of different color in Fig. 1 to clearly show the structure of dendritic cell cluster. A concrete interpretations of dendritic cell cluster has described in the author's previous paper [30]. The transmitted electric field of dendritic cell cluster metasurface is shown in Fig. 1d. Cross-polarized transmitted plane wave $E_x$ is also obtained; the angle of refraction is 75°. The intensity of transmitted electric field in the dendritic cluster metasurface is greater than that in the metasurface composed of a single type of dendritic structure shown in Fig. 1b. The simulation result suggests that dendritic metasurface has the ability of manipulating light in cross-polarized conversion and negative refraction. The intensity of cross-polarized plane wave in anomalous refraction direction is even greater when dendritic metasurface is composed of many different types of dendritic cell.

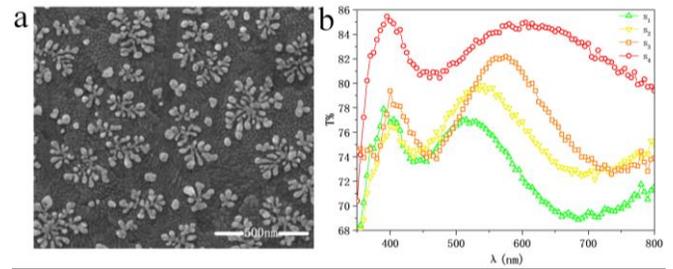

**Fig. 2.** (a)SEM photograph of silver dendritic cell cluster metasurface; (b) The transmission spectrum of dendritic cell cluster metasurface.

The dendritic cell-cluster metasurface is prepared with the bottom-up electrochemistry deposition. The dendritic metasurfaces sample consists of three layers: the bottom layer is a substrate which is made up of tin indium oxide electroconductive glass; the interface layer is a 2D evenly distributed individual silver dendritic cell (Fig. 2a); whereas the top layer is the oxidation resistant coating. This provides a full coverage of the transmission phase range. As shown in Fig. 2a, the dendritic structure is zoomed in at multiple of $2 \times 10^5$ in scanning electron microscope. The dimension of a single dendritic unit is approximately 200–300 nm, uniformly distributed onto the substrate surface. Several dendritic units coupled nearby with diverse size and branches to form a cluster in the dendritic metasurface. For convenience in experimental measurement, the overall dimension of the dendritic metasurface is 1 × 1.3 cm. A larger sample will be obtained with corresponding size substrate. The transmitted spectra of the four types of dendritic metasurface (s1, s2, s3 and s4) are shown in Fig. 2b. In addition to the silver intrinsic absorption peak at approximately 400 nm, high transmission peaks are also found in the wavelength region of 510-530 nm (s1), 530-555 nm (s2), 555-580 nm (s3) and 600–620 nm (s4). Sample responses in different visible wavelength ranges were prepared by properly tuning the deposition condition. The light source with tunable wavelength will be employed to test the dendritic cell-cluster metasurface.

The optical transmission of the dendritic metasurface is measured using a tunable broadband source, shown in Fig. 3a. A xenon light coupled with a visible–near-infrared monochromator is used as the tunable light source (with a wavelength range of 300–2,000 nm). The generated plane wave from the monochromator is circular polarized. In this experiment, the wavelength of incident light varying from 490 to 640 nm with a step length of 10 nm was set. D is the aperture diaphragm. To control the polarization angle of the incident light, polarizer P1 is used to convert the circular polarized incident light to linear polarized light $E_y$. The sample of dendritic metasurface to be measured which lay perpendicularly to the incident light. Then, the linear polarized incident light $E_y$ is focused on the surface of the dendritic metasurface using a planoconvex L (with a focal length of 50 mm). The polarizers P2 and P3 behind sample are used to detect the polarized direction of the transmitted light. The transmitted light spot is received by a semitransparent white plate. Real-time observation of the phenomena of all transmitted light is achieved using a charge-coupled device (CCD) camera. All elements of the experiment are placed on a self-balancing optical table; the measuring process is operated in an optical darkroom.

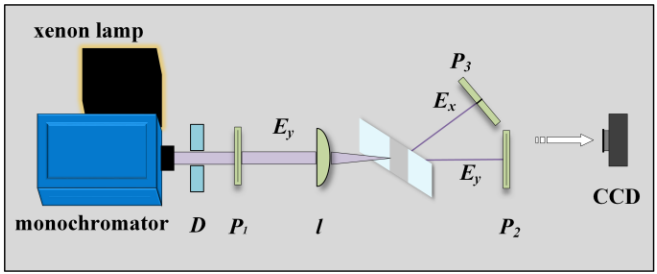

**Fig. 3.** Schematic of experimental setup used to measure the transmitted light through the metasurface. D is the aperture diaphragm, P1, P2 and P3 are polarizers, L is a planoconvex and CCD is a charge-coupled device camera.

The wavelength of incident light is increased from 490 nm to 640 nm with a uniform speed; meanwhile, the phenomena of the transmitted light are recorded using a CCD camera. In this manner, the video of transmitted light through the dendritic metasurface with varied incidence wavelength was obtained. As shown in the obtained video (Supplementary Materials), two optical spots are observed on the white plate when the wavelength of incident light is within the range of resonance wavelength. The bright one at the center of the white plate reveals that the normal transmitted light is perpendicular to the sample interface; the other one is with low brightness beside the center of the anomalous transmitted light. These two optical spots reveal that the dendritic metasurface can deflect light propagation into anomalous refraction channels. The polarization analyzers P2 and P3 are employed to measure the polarization angle of the normal and anomalous transmitted light. The measuring result show that the polarization angle of the normal transmitted light is the same as the incident light. The polarized direction of the anomalous transmitted light is perpendicular to the polarized direction of the incident light, i.e., the normal transmitted light is co-polarized, and the anomalous transmitted light is cross-polarized. In the nonresonance wavelength, only one single light spot at the center of the white plate is observed. The measurement result of the polarization analyzer reveals that the transmitted light is a co-polarized light.

Referring to the transmitted spectra curve of dendritic metasurface in Fig. 2, when the wavelength of the incident light occurs at the resonance wavelength of the sample, in addition to the normal co-polarized transmitted light, a cross-polarized transmitted light is obtained. When the wavelength of incident light and the resonance wavelength of the sample are inconsistent, the transmitted light is co-polarized along the original propagation path. Further, to illustrate the connection between the resonance wavelength of dendritic metasurface and the operation wavelength of cross-polarization conversion, the responses of the four samples are measured. The measured results are shown in Fig. 4. The cross-polarized transmitted light of sample $s_1$ operating at a wavelength of 550 nm is obtained (Fig. 4a). The response wavelength of the measuring sample is approximately 520 nm (Fig. 2b). The cross-polarized transmitted light of samples $s_2$, $s_3$ and $s_4$ are respectively shown in Fig. 4b, 4c and 4d. In addition, the cross-polarized transmitted light with lower intensity than the co-polarized light. The cross-polarized light intensity of sample $s_3$ is highest in all measured samples as shown in Fig. 4 and the transmission of sample $s_3$ is highest as shown in Fig. 2b. In summary, the resonance wavelength of dendritic metasurface and the operation wavelength of cross-polarization conversion using the dendritic metasurface is corresponding. The response wavelength of the dendritic metasurface is controlled during the preparation. The cross-polarized transmitted light operates in a different wavelength is obtained using the dendritic metasurface response in a corresponding wavelength.

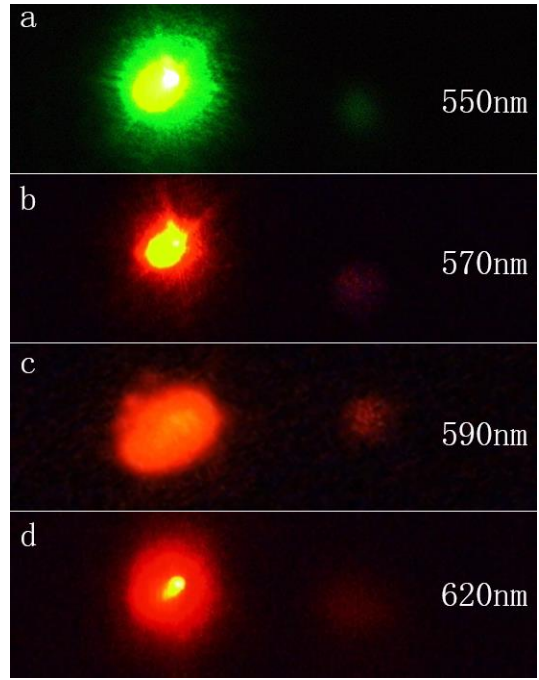

**Fig. 4.** Photograph of the experimental co- and cross-polarization transmission of (a) $s_1$ within 550 nm wavelength; (b) $s_2$ within 570 nm wavelength; (c) $s_3$ within 590 nm wavelength (d) $s_4$ within 590 nm wavelength incident.

The refraction angle of the co-polarized transmitted light is 0°; the refraction angle of the cross-polarized transmitted light is approximately 60°. The refractive angle of the cross-polarized transmitted light is approximately 75° in the simulation. A certain bias exists between the experimental and the simulation results mainly because of the prepared dendritic metasurface sample in the chemical method does not completely agree with the model in the simulation process. The simulated and experimental results revealed that the dendritic cell-cluster metasurface achieves cross-polarization conversion of the linearly polarized incident light in the resonant frequency of the sample. In the improved experimental design and technology of preparation, the conversion efficiency of cross-polarization is expectedly enhanced. The dendritic cell-cluster metasurface prepared using the bottom-up electrochemical deposition method has great potential in the light manipulative applications.

Here, a dendritic cell cluster cross-polarized conversion metasurface operating in the visible light with the transmission mode was proposed. A silver dendritic metasurface in the visible light was prepared with the electrochemical deposition method based on the bottom-up concept. Numerical simulation and experiments confirmed that when the wavelength of the incident light coincide with the sample resonant band, co- and cross-polarization transmission were obtained after the linear polarized incident light perpendicularly passed through the silver dendritic metasurface, the co-polarization light was perpendicular to the interface, and a tilted cross-polarization was emitted. Further improving the preparation process of the dendritic metasurface will obtain higher conversion efficiency of the dendritic metasurface, which opened a new chapter in the practical application of light manipulation.

.

**Funding.** National Natural Science Foundation of China (51272215, 11674267) and the National Key Scientific Program of China (under project No. 2012CB921503).

**Acknowledgment**.